\documentclass[%
 aip,
 amsmath,amssymb,
 reprint,%
]{revtex4-1}

\usepackage{graphicx}% Include figure files
\usepackage{dcolumn}% Align table columns on decimal point
\usepackage{bm}% bold math
\usepackage[utf8]{inputenc}
\usepackage[T1]{fontenc}
\usepackage{mathptmx}
\usepackage{etoolbox}
\usepackage{xcolor}
\usepackage{color}

\usepackage[normalem]{ulem}
\makeatletter
\def\@email#1#2{%
 \endgroup
 \patchcmd{\titleblock@produce}
  {\frontmatter@RRAPformat}
  {\frontmatter@RRAPformat{\produce@RRAP{*#1\href{mailto:#2}{#2}}}\frontmatter@RRAPformat}
  {}{}
}%
\makeatother
\begin{document}

\preprint{AIP/123-QED}

\title{Fully reversible magnetoelectric voltage controlled THz polarization rotation in magnetostrictive spintronic emitters on PMN-PT}
% Force line breaks with \\
\author{G.Lezier}%
\affiliation{ 
Univ. Lille, CNRS, Centrale Lille, Univ. Polytechnique Hauts-de-France, UMR 8520-IEMN,F-59000 Lille,France}%
\author{P. Koleják}
\affiliation{ 
Univ. Lille, CNRS, Centrale Lille, Univ. Polytechnique Hauts-de-France, UMR 8520-IEMN,F-59000 Lille,France}%
\affiliation{Technical University of Ostrava, IT4Innovations \& Faculty of Materials Science and Technology, 708 00 Ostrava - Poruba, Czech Republic}

\author{J-F. Lampin}%
\affiliation{ 
Univ. Lille, CNRS, Centrale Lille, Univ. Polytechnique Hauts-de-France, UMR 8520-IEMN,F-59000 Lille,France}%

\author{K. Postava}

\affiliation{Technical University of Ostrava, IT4Innovations \& Faculty of Materials Science and Technology, 708 00 Ostrava - Poruba, Czech Republic}%
\author{M. Vanwolleghem}%
\author{N. Tiercelin}
 \email{nicolas.tiercelin@iemn.fr}

\affiliation{ 
Univ. Lille, CNRS, Centrale Lille, Univ. Polytechnique Hauts-de-France, UMR 8520-IEMN,F-59000 Lille,France}%

\date{\today}% It is always \today, today,

\begin{abstract}
THz polarization control upon generation is a crucially missing functionality. THz spintronic emitters based on the inverse spin Hall effect allow for this by the strict implicit orthogonality between their magnetization state and the emitted polarization. This control was up till now only demonstrated using cumbersome external magnetic field biasing to impose a polarization direction.
We present here for the first time an efficient voltage control of the polarization state of terahertz spintronic emitters. Using a ferromagnetic spin pumping multilayer exhibiting simultaneously strong uniaxial magnetic anisotropy and magnetostriction in a crossed configuration, an emitter is achieved where in principle the stable magnetization direction can be fully and reversibly controlled over a 90$^\circ$ angle span only by an electric voltage. To achieve this, an engineered rare-earth based ferromagnetic multilayer is deposited on a piezoelectric $\textrm{(1-x)[Pb(Mg}_\textrm{0.33}\textrm{Nb}_\textrm{0.66})\textrm{O}_\textrm{3}\textrm{]-x[PbTiO}_\textrm{3}\textrm{]}$ (PMN-PT) substrate. We demonstrate experimentally a reversible 70$^\circ$ THz polarization rotation by sweeping the substrate voltage over 400V. This demonstration allows for a fully THz polarization controlled ISHE spintronic terahertz emitter not needing any control of the magnetic bias.

\end{abstract}

\maketitle

Spintronic Inverse Spin Hall Effect terahertz emitters (ISHE STE) have in a matter of just a few years burst onto the scene as a novel promising and highly efficient paradigm for pulsed terahertz generation \cite{Kampfrath2013,Seifert2016,Papaioannou2018}. They compete with or even outperform standard schemes based on nonlinear optical rectification or ultrafast semiconductor-based photoswitches, both in terms of emitted power and especially bandwidth, while adding a unique direct access to the polarization direction of the emitted terahertz pulse by control of the magnetization direction of the spin pumping layer \cite{Hibberd2019}. The physics of the ISHE inherently imply the emitted THz pulse to be strictly perpendicular to the magnetization of the spin pumping layer. This is a particularly interesting feature enabling for instance a compact and efficient source for time-domain terahertz ellipsometry \cite{nagashima2001,chen2018}. Moreover, fast modulation of the magnetization direction opens the way for an efficient modulation format for THz wireless communications and could also find use in low-noise modulation ellipsometric THz spectroscopy \cite{Gueckstock2021}. While optimizations of the THz power have been reported\cite{Nenno2019,Torosyan2018}, the control of the THz polarization state has been limited to cumbersome implementations using perpendicularly crossed or mechanically rotating external magnetic fields\cite{KongAOM2019,Yang2016}. Recently, we demonstrated a ISHE STE with a spin pumping layer with built-in uniaxial magnetic anisotropy \cite{KolejakLezier2021}. This achieves full 2$\pi$-radians control of the polarization by setting only a single hard-axis external magnetic field. Although efficient, this method still requires an electromagnet and continuous currents to maintain the desired polarization. In the present work, we improve on this scheme. The intermetallic exchange coupled CoFe(5\AA)/TbCo2(8\AA)/CoFe(5\AA) multilayer (ECML) designed for the purpose of uniaxial magnetic anisotropy, is also known for its strong magnetoelastic properties \cite{DebrayLudwig2004, Dusch2013}. As such, the magnetic anisotropy can be fully controlled by applying an appropriate in-plane tensile strain. Using a piezoelectric substrate, the magnetization direction in the spin pumping layer can therefore be set solely by an electrical voltage. Owing to this control of the magnetic anisotropy, one can theoretically reach a 90° polarization rotation. Recently, Cheng et al. presented the study of a magnetoelectric effect in spintronic emitters using an iron-cobalt-boron alloy \cite{ChengHuang2021}. They could only obtain a 30° polarization rotation plagued moreover by an important loss of THz signal, and exhibiting very limited reversibility. We hereby experimentally demonstrate for the first time a scheme achieving a fully reversible voltage controlled coherent terahertz polarization rotation using a strain mediated magnetoelectric effect. To this aim, an exchange-coupled Pt/FeCo/TbCo2 /FeCo/W emitter deposited onto an electro-active (1-x)[Pb(Mg0.33 Nb0.66 )O3 ]-x[PbTiO3 ] (PMN-PT) substrate \cite{WangLuo2007} . 

We first demonstrate how by proper design the interplay of the magnetostriction and the uniaxial magnetic anisotropy of the ECML theoretically allows for a 90◦ rotation of the magnetization by a voltage control. The fabrication process of our emitters is then described.  Their THz emission is characterized using an in-house customized terahertz time-domain spectroscopy setup, allowing a full vectorial measurement of the generated polarization. Finally, a phenomenological macrospin model that takes into account both the Zeeman energy term, the uniaxial magnetic anisotropy energy and the magneto-elastic energy, correctly reproduces the experimentally measured THz polarization rotation sweep and helps to understand the current limitations of the polarization angle span. 

\begin{figure}
    \includegraphics[width=\columnwidth]{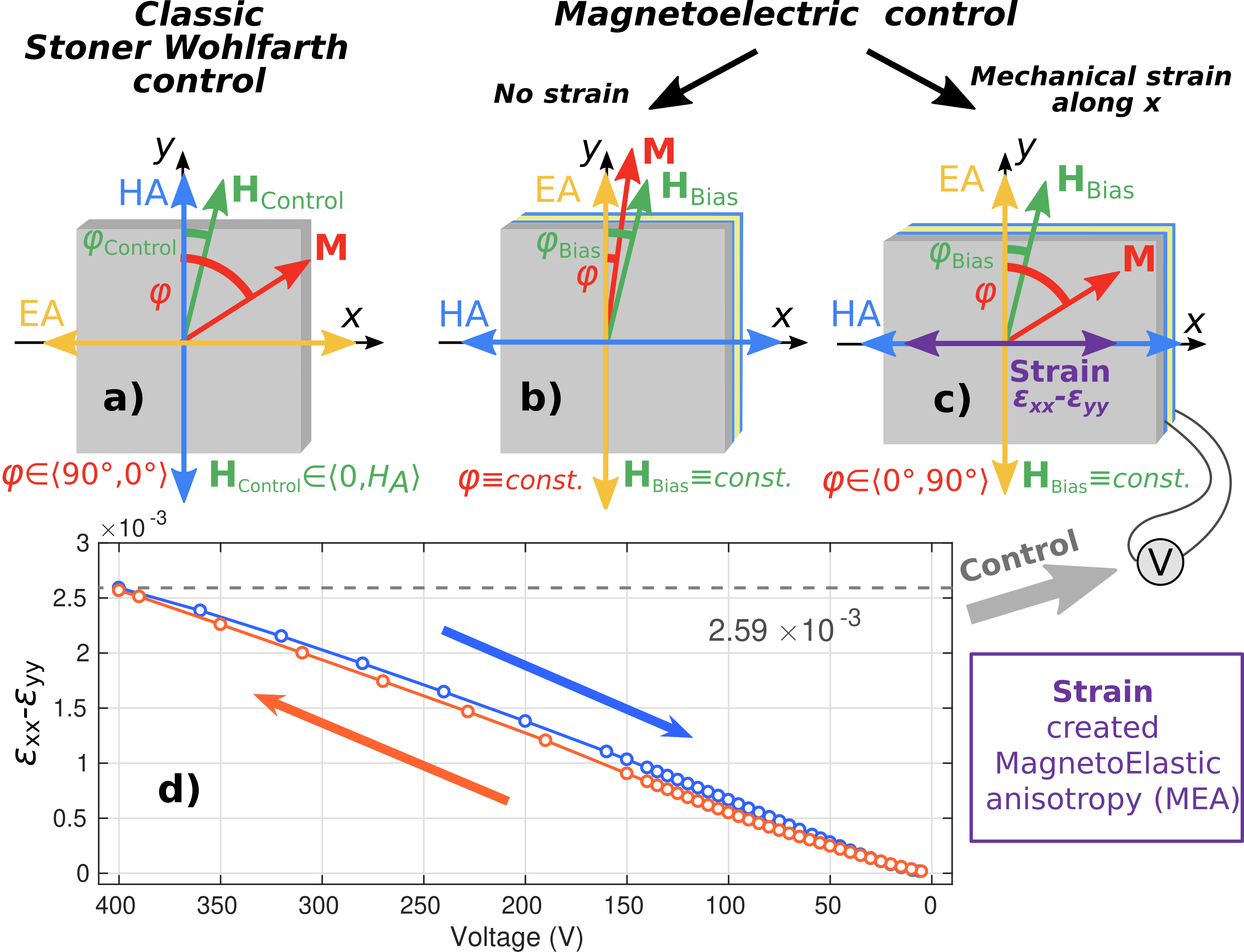}
    \caption{a) Stoner-Wolhfarth control of polarization; b) and c) Magnetoelecric control; d)  $\varepsilon_{xx}-\varepsilon_{yy}$ PMN-PT anisotropic deformation as a function of the applied voltage while cycling from 400V to 0V and back to 400V.}
    \label{fig:strain}
\end{figure}

As stated earlier, we recently demonstrated the implementation of an ISHE STE based on a ferromagnetic layer possessing a uniaxial magnetic anisotropy \cite{KolejakLezier2021}. As schematized on figure \ref{fig:strain}a, the polarization control mechanism is based on the Stoner-Wohlfarth-like reorientation of the magnetization M in a uniaxial anisotropic layer submitted to a $\textrm{H}_\textrm{control}$ magnetic field along the orthogonal direction to the easy axis (EA). The magnetization angle is determined by the competition between the Zeeman energy due to the applied field and the anisotropy energy. Varying $\textrm{H}_\textrm{control}$ between 0 and the value of the anisotropy field $\textrm{H}_\textrm{A}$ allows for a rotation of almost 90° for the magnetization and thus polarization of the emitted THz signal. For symmetry reasons, a slight tilt of the magnetic field must be set to prevent the creation of magnetic domains\cite{KlimovTiercelin2006} and ensure a coherent rotation of M. For the magnetoelectric control represented on figure \ref{fig:strain}b,c, we can exploit the fact that a strong anisotropy can be induced by straining the ferromagnetic layer if it is also magnetostrictive. The added magnetoelastic anisotropy (MEA) energy competes with the two other terms and allows for a voltage control of M if the strain is generated by an electroactive substrate. Without strain, and in the configuration depicted in figure\ref{fig:strain}b, M lies close to the vertical axis. A magnetic bias is also added to ensure no magnetic domains are created. Upon application of a tensile strain in the x direction (compressive in the y direction), a MEA is created along x and M tends to align with it (Fig. \ref{fig:strain}c). For a high enough strain, this anisotropy overcomes the uni-axial anisotropy and bias, and thus theoretically, the maximum angle span for the polarization rotation is 90°. For this principle to work properly, the deformation has to be strongly anisotropic, since the associated magnetoelastic energy is proportional to $b^{\gamma,2}\left( \varepsilon_{xx}-\varepsilon_{yy} \right)$ where $\varepsilon_{xx}$ and $\varepsilon_{yy}$ are the relative displacements and $b^{\gamma,2}$ is the magnetoelastic coefficient of the layer. Suitable substrates are single crystal PMN-PT in the <011> cut as they possess $\textrm{d}_{31}$ and $\textrm{d}_{32}$ piezoelectric coefficients with opposite signs\cite{Dusch2013}.\\
The chosen PMN-PT substrates were commercial <011>-cut single crystal 10mm $\times$ 10mm $\times$ 0.5mm squares of the 'X2A' reference provided by the TRS technologies company \cite{trstechnologies}. Firstly, the anisotropic piezoelectric in-plane deformation of the PMN-PT was measured as a function of the applied voltage with a standard double strain-gauge technique\cite{WuZhao2011}. With this method, we were able to detect the sign of the in-plane deformations and make sure the extension of the substrate is in the horizontal direction x. The anisotropic deformation value $\varepsilon_{xx}-\varepsilon_{yy}$ is shown on Figure \ref{fig:strain}d and is suitable for the magnetoelectric control.

The previously optimised W/FeCo/TbCo$_2$/FeCo/Pt ISHE emitters are grown on those substrates. Prior to deposition, both sides were mechanically polished down to a surface roughness of less that 1nm over a $\textrm{2}\mu \textrm{m}$ by $\textrm{2}\mu \textrm{m}$ area. The cleaning procedure involved 5 minutes ultrasonication steps in successive baths of dichloromethane, acetone, isopropyl alcohol and deionized water. The W(2nm)/CoFe(5\AA)/TbCo$_2$(8\AA)/CoFe(5\AA)/Pt(2nm) spintronic emitters were deposited by RF-Diode sputtering in a LEYBOLD Z550 equipment. The typical base vacuum in the deposition chamber is $3\cdot10^{-7}$mbar. During deposition, the substrates are placed in an in-plane field with an approximate strength of 80 kA/m to imprint an in-plane magnetic anisotropy along the Y direction. The individual layers are sputtered from circular 4 inches targets. The TbCo$_2$ and FeCo alloys are obtained from composite targets. W, Pt and FeCo are deposited with a RF power P$_{\mathrm{RF}}$ of 440W under a pure Argon atmosphere at 2$\cdot$10$^{-3}$~mbar. TbCo$_2$ is deposited with a RF power P$_{\mathrm{RF}}$ of 250W at 10$^{-3}$~mbar. To ensure a high precision in the thicknesses, the deposition was carried on a rotatory turn-table substrate holder in an oscillation mode with a calibrated rotation speed. The CoFe/TbCo$_2$/CoFe tri-layer acts as an exchange-coupled multilayer and the 5d metals Pt and W are known to provide strong spin-orbit coupling leading to record spin Hall angles \cite{Seifert2016}. On the backside of the substrate, an additional 2nm platinum layer is sputtered to serve as an electrode to apply the driving voltage to the PMN-PT.\\

\begin{figure}
    \includegraphics[width=\columnwidth]{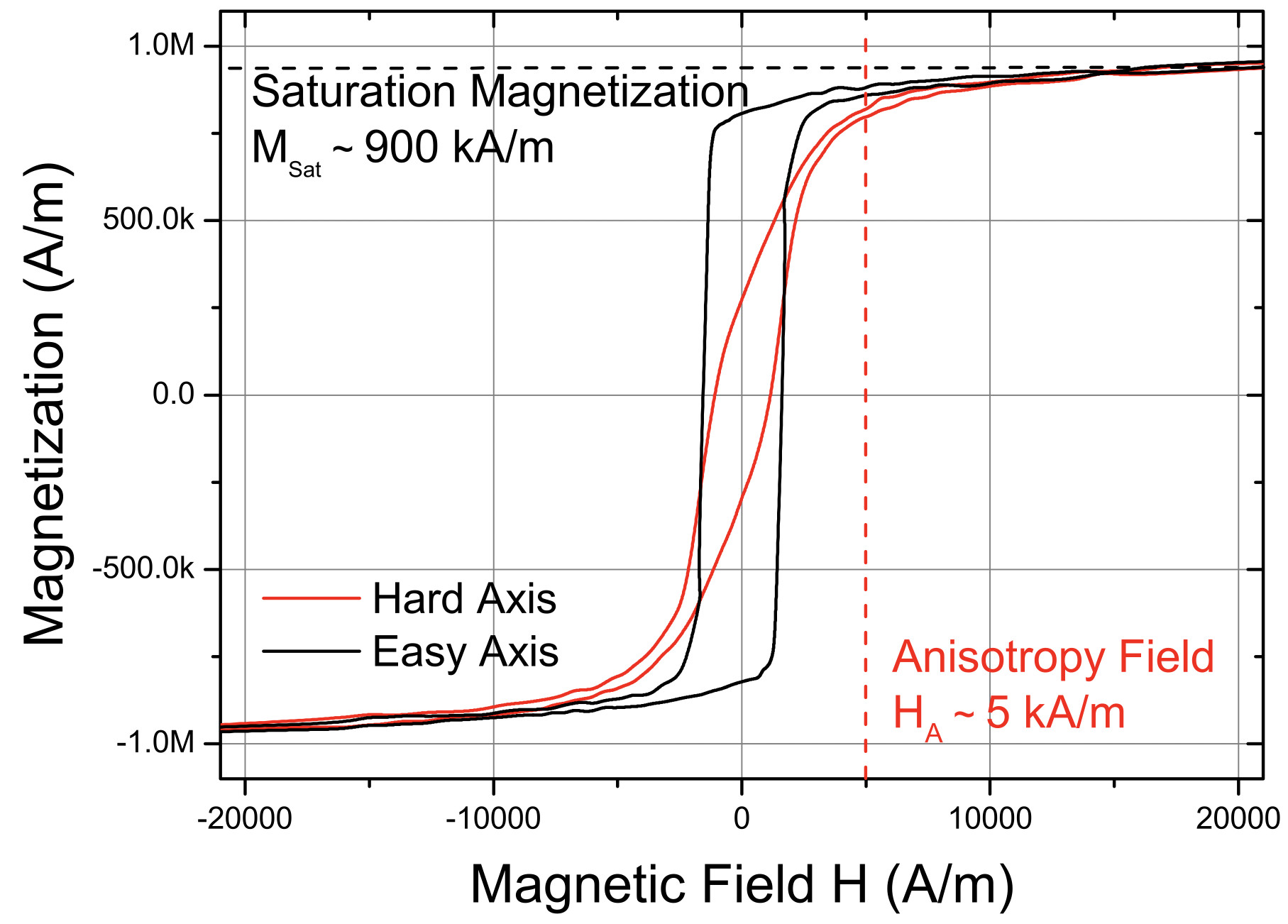}
    \caption{VSM magnetization cycles for the spintronic emitter along the easy and hard axes. 
    }
    \label{fig:VSM}
\end{figure}

\begin{figure}
    \includegraphics[width=\columnwidth]{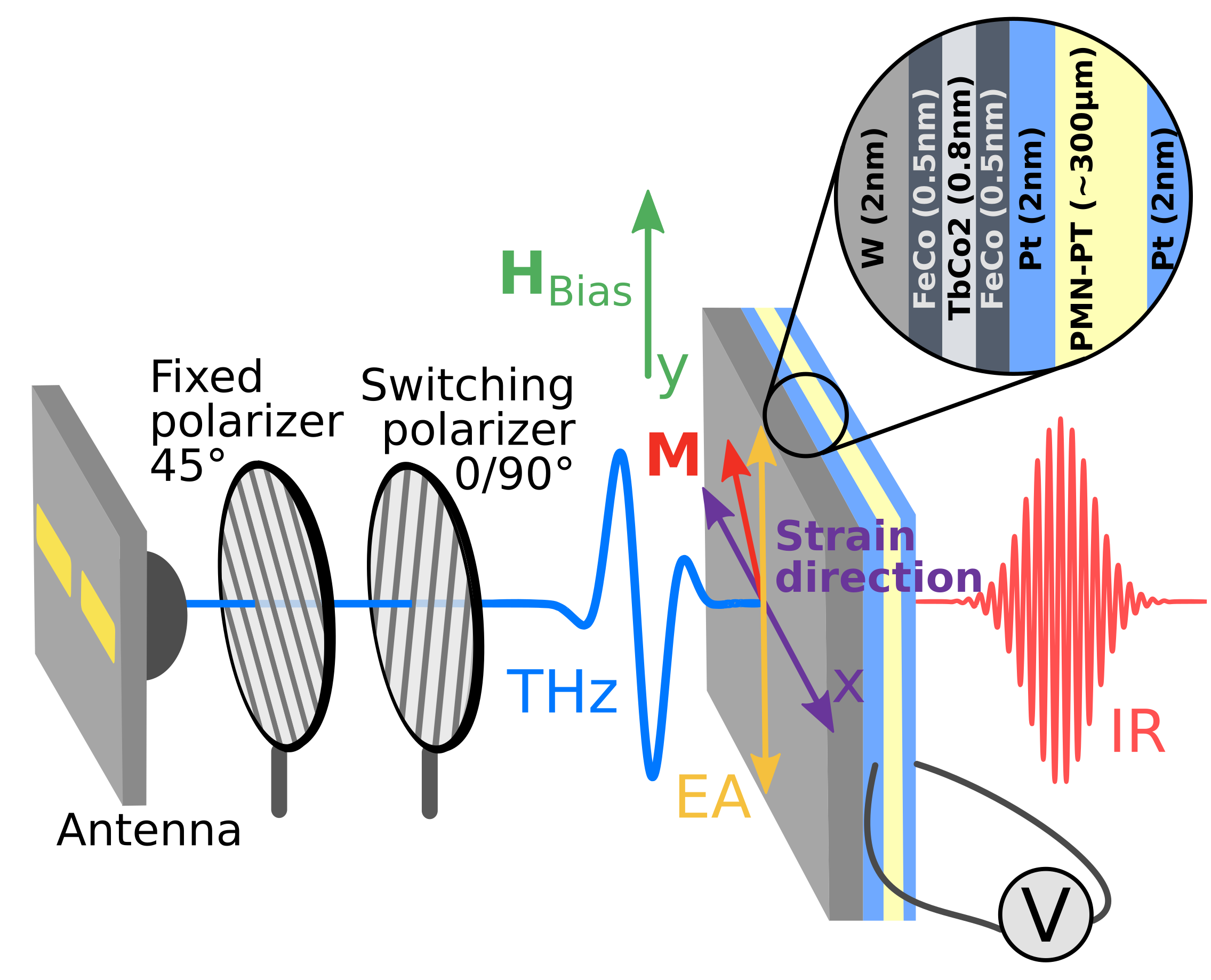}
    \caption{Layer composition of the strain-mediated ISHE STE and polarimetric measurements scheme.}
    \label{fig:setup}
\end{figure}

The magnetization behaviour is characterized with a vibrating sample magnetometer (VSM). With this kind of magnetic stack, a magnetic easy axis (EA) is usually created in the direction of the applied field during the deposition. In the present case, while the uni-axial anisotropy is present, we found a slight offset of 7° with respect to the deposition field. While the origin of this effect is not clear, it is certainly due to the unusual nature of the PMN-PT substrate. The resulting anisotropy direction can be altered by the crystalline structure of the <011>-cut PMN-PT or residual anisotropic stress. Figure~\ref{fig:VSM} shows the magnetization loops along the easy and hard axis of the layer. They confirm the anisotropic nature of the layer, and the corresponding anisotropy field $\textrm{H}_\textrm{A}$ is estimated to be approximately equal to 5 kA/m. The saturation magnetization is close to 900kA/m. A schematic view of the emitter geometry is shown on Figure~\ref{fig:setup}.\\

\begin{figure}
    \includegraphics[trim=50 0 50 0,clip,width=\columnwidth]{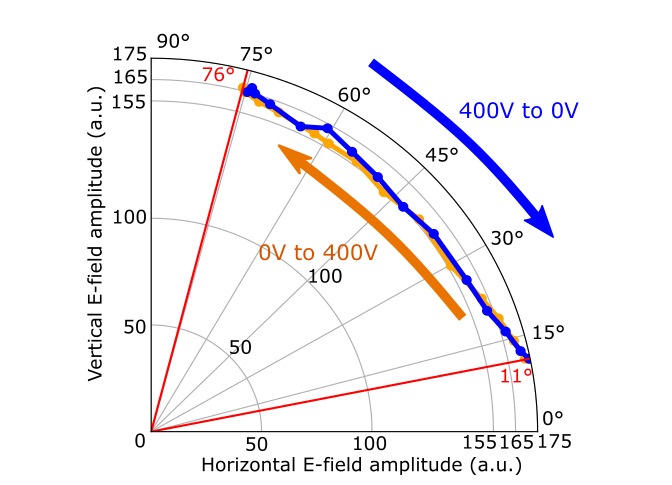}
    \includegraphics[trim=320 0 350 0, clip,width=\columnwidth]{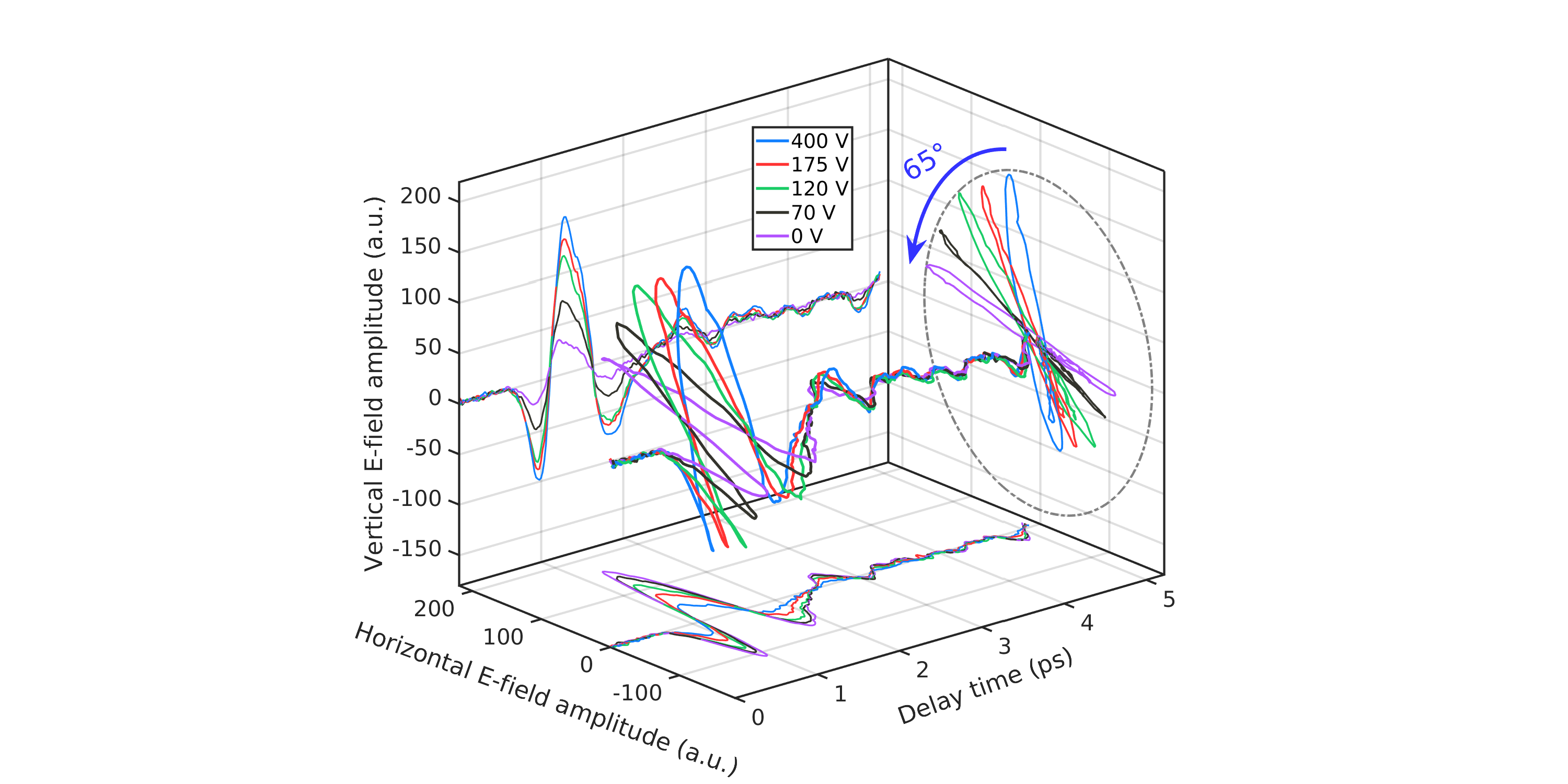}
    \caption{Top : Emitted electric field polarization and amplitude as a function of the applied driving voltage for an applied bias of 1.5kA/m. Bottom : Temporal TDS spectra for 400V, 175V, 120V, 70V and 0V. 
   }
    \label{fig:rotation}
\end{figure}

The THz emission was characterized on a customized terahertz time-domain spectroscopy (TDS) setup. The ISHE-mediated terahertz emission is generated by pumping the sample with femtosecond pulses from a Ti:sapphire laser oscillator (80MHz repetition rate, center wavelength 800nm and 100fs pulse duration). The E-field of the emitted terahertz pulse is measured by sampling the response of a photoconductive Auston switch that is probed by a split-off fraction of the femtosecond infrared pulse by a delay line. In order to measure the horizontal and vertical components of the E-field and deduce the polarization angle, two wire-grid polarizers were inserted in the THz path. Since the receiving antenna is horizontal, a switching polarizer is used to select the horizontal or vertical projections of the electric field to be sampled, while a fixed polarizer ensures that a signal is detected with the same amplitude regardless of the projection as shown in Figure \ref{fig:setup}. The emitter is maintained by a flexible adhesive on its holder, so that its free deformation is allowed. Wires are bonded to the front and back electrodes with silver conductive epoxy and connected to a SHR High Voltage Source. Firstly, the TDS setup delay line was set to obtain the maximum signal amplitude. Preliminary voltage cycles showed a maximum polarization rotation towards the vertical direction for voltages of about 400V. The E-field components were then measured while cycling the applied voltage on the PMN-PT substrate from the previously determined saturation of 400V to 0V. The tilted magnetic bias $\textrm{H}_\textrm{Bias}$ ensures that the formation of domains is minimal in the magnetic film. To that end, a 1.5kA/m magnetic field is applied, with an arbitrarily chosen 10$\circ$ tilt from the vertical direction to be superior to the 7$\circ$ angle of anisotropy. The deducted polarization angle shown on Figure 4 clearly shows the controlled rotation of the electric field polarization over a span of more than 65 degrees. The emission behaves as expected: under no applied field, the emitted signal has a horizontal polarization since magnetization lies vertically along the EA. Upon a voltage increase, the <011> PMN-PT generates a tensile strain in which induces an anisotropy along the horizontal direction forcing magnetization to rotate to its new equilibrium position. A voltage decrease lowers the strain induced magnetic anisotropy and the magnetization quasi-reversibly rotates back to its equilibrium position at 0V. From that measurement, we can determine the polarization angle as a function of the applied voltage. The time domain spectroscopy measurements shown in Figure \ref{fig:rotation} at fixed voltage values confirm the rotation of the polarization. Along with the rotation, a slight ellipticity appears for large values of applied voltage. Several hypotheses can be drawn to explain this : (i) part of the generated THz signal is received after reflection off the (lossy) PMN-PT substrate that it itself possesses an important dielectric anisotropy as expected from its orthorhombic crystal structure. This will result in a phase lag between the reflection coefficients of the horizontal and vertical field components. Moreover, the anisotropy of PMN-PT is increasing with voltage in view of its strong piezolectricity\cite{WangLuo2007, KollaYushin1998}; (ii) it is reasonable to expect a slight deviation of the uniformity of the strain produced by the substrate. This causes the magnetoelastic anisotropy to be slightly "spread" over the generated THz beam resulting in a spread rotation and thus an ellipticity. In any case, further detailed modeling and analysis is needed to elucidate this ellipticity which seems to be limited to $\approx$ 10 in the worst case (at 400V).

As shown on Figure \ref{fig:rotation}, the polarization angle span we obtained is of about 65°, whereas the control via the magnetoelastic anisotropy can theoretically allow a 90° span. But several geometrical factors have to be accounted for, and in particular, other contributions to the magnetic energy of the system, namely the intrinsic uni-axial anisotropy of the layer and the bias that is applied to limit domain formation during the rotation process. 

\begin{figure}
    \includegraphics[width=\columnwidth]{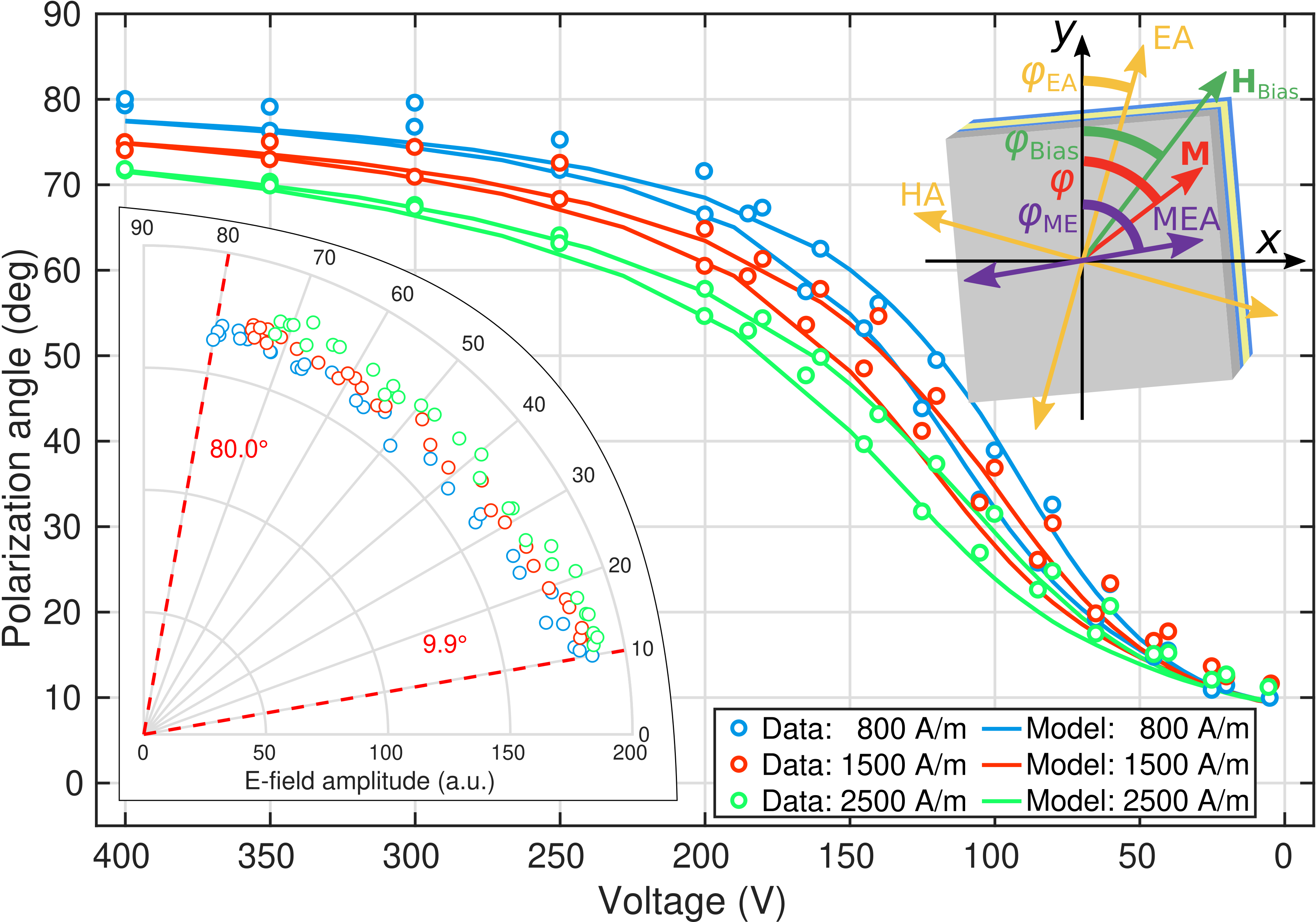}
    \caption{Comparison between experimental data and model for the polarization angle as a function of voltage while cycling from 400V to 0V and back to 400V, for three values of the applied bias in the  $\varphi_\textrm{Bias}$=10° direction. Top right inset: Angle configuration. Bottom left inset: experimental measurements polar plot.}
    \label{fig:model}
\end{figure}

To understand the influence of these parameters, a macrospin model was used to calculate the angle of magnetization when the emitter is submitted to the anisotropic strain generated by the PMN-PT when voltage is applied. We consider here that the magnetization coherently rotates in the plane of the film, which is a common assumption in Stoner-Wohlfarth-like anisotropic layers submitted to a magnetic bias. The anisotropic deformation value $\varepsilon_{xx}-\varepsilon_{yy}$ previously measured is introduced in the algorithm. For each voltage value, it will determine the angular position of the magnetization following the minimum of magnetic energy which is the sum of the Zeeman interaction energy, the anisotropy energy due to the intrinsic EA of the film, and the magneto-elastic energy: 
\begin{multline}
    E_m=-\mu_0\mathrm{M}_\mathrm{Sat}\mathrm{H}_\mathrm{Bias}\cos\left( \varphi-\varphi_\mathrm{Bias} \right)\\
    -\frac{1}{2}\mu_0\mathrm{M}_\mathrm{Sat}\mathrm{H}_\mathrm{A}\cos^2\left( \varphi-\varphi_\mathrm{EA} \right)\\
    +b^{\gamma,2}\left( \varepsilon_{xx}-\varepsilon_{yy} \right)\cos^2\left( \varphi-\varphi_\mathrm{ME} \right)
\end{multline}
 where $\mu_0$ is the vacuum permeability, $\textrm{M}_\textrm{Sat}$ the value of the saturation magnetization, $\textrm{H}_\textrm{Bias}$ the strength of the magnetic bias field, $\varphi$ the angular position of magnetization to be determined, $\varphi_\mathrm{Bias}$ the angle of the applied magnetic field, $\textrm{H}_\textrm{A}$ the value of the anisotropy field, $\varphi_\mathrm{EA}$ the direction of the intrinsic EA, $\textrm{b}^{\gamma,2}$ the magnetoelastic coefficient of the ferromagnetic film, and $\varphi_\mathrm{ME}$ the angle of the anisotropy created by the PMN-PT deformation. For the calculations, all the considered angles are taken with respect to the vertical polarization component in the (X,Y) referential given by the polarizers, as shown on the inset in Figure \ref{fig:model}. Owing to that $\varphi_\mathrm{Bias}$, $\varphi_\mathrm{EA}$ and $\varphi_\mathrm{ME}$ can be determined using TDS polarimetric measurements, since the emitted polarization is perpendicular to magnetization : Cutting off the voltage and applying a large magnetic bias field of more than 15kA/m will force the magnetization along $\varphi_\mathrm{Bias}$. Then it will align along $\varphi_\mathrm{EA}$ when cutting off the bias field. Finally, a 500V voltage is applied to the PMN-PT in order to create a large anisotropy in the $\varphi_\mathrm{ME}$ direction. In the chosen configuration we find $\varphi_\mathrm{Bias}=10^\circ$, $\varphi_\mathrm{EA}=9^\circ$, $\varphi_\mathrm{ME}=85^\circ$. From the VSM measurements, we find $\textrm{M}_\textrm{Sat}\textrm{=900kA/m}$ and $\textrm{H}_\textrm{A}\textrm{=5kA/m}$. 
 
 Figure \ref{fig:model} shows the results of the calculations and the comparison with experimental measurements for three different values of the bias field. Both simulations and experiments agree very well, assuming a magnetoelastic coefficient $\textrm{b}^{\gamma,2}$ equal to -2.6MPa. Given the 2nm thickness of the ferromagnetic layer, it was not possible to determine its magnetoelastic coefficient, but such a value is consistent with Terbium-transition metal based multilayer systems \cite{JuraszekGrenier2006,BenYoussefTiercelin2002}. The hysteresis in the cycles is due to the hysteresis of the strain in the PMN-PT shown on Figure \ref{fig:strain}.\\
 
From Figure \ref{fig:model}, we can safely deduce that as expected, the magnetization M of the ferromagnetic layer, and thus the polarization of the emitted THz signal, is indeed governed by the deformation of the PMN-PT substrate upon application of the driving voltage: the strain induced anisotropy competes with the bias Zeeman energy and the original EA anisotropy. For low voltages, the magnetization angle is set by the competition between the EA and the bias which is close to 10°. Upon increase of the voltage, M rotates towards the $\varphi_\textrm{ME}$ direction. Here, we can see the influence of the bias field : as it increases, it makes it harder for M to rotate towards $\varphi_\textrm{ME}$, reducing the angle span. With $\textrm{H}_\textrm{Bias}$ =1500A/m the angle span is 65°. Reducing $\textrm{H}_\textrm{Bias}$ down to 800A/m increases the span to 70°, while for $\textrm{H}_\textrm{Bias}\textrm{=2500A/m}$ this span is only 60°. Nevertheless, there is a trade-off here since reducing the bias too much results in a decrease of the total strength of the THz strength (see Fig.~\ref{fig:model} inset). This is likely an indication of magnetic domain formation preventing the rotation of M to be purely homogeneous. At last, we see that in order to increase the span, the EA and the direction of  $\textrm{H}_\textrm{Bias}$ have to be as perpendicular as possible from the Magneto-Elastic Anisotropy (MEA) orientation. Since the MEA direction is linked to the PMN-PT substrate, the EA has to be defined accordingly. As it was stated earlier, we noted that after the sputter deposition of our STE under a magnetic field, the EA has a 7°-degree offset compared to the desired direction, which is consistent with our polarization angle span. Further studies on the origin of this offset have to be led, but magnetic field annealing might also improve this characteristic and help to reach a close to 90° polarization span. \\

Overall, we presented here the first demonstration of a reversible magnetoelectric control of THz polarization rotation that paves the way to a new class of voltage-controlled THz devices. Thanks to the use of an anisotropic and magnetoelastic ferromagnetic layer in a spintronic emitter deposited on a PMN-PT substrate, our experimental device was able to show a reversible and repeatable control of the THz polarization with a 70° span for 400V. Our macrospin model confirmed the control mechanism, based on the creation of a magnetoelastic anisotropy, and gave insight on how to improve the device. This demonstration makes the intrinsic polarization control allowed by ISHE emitters more practical by removing the need for precise magnetic bias control. Only a constant external field is needed to prevent domain formation during rotation.

\begin{acknowledgments}
The authors acknowledge financial support from the Horizon 2020 Framework Programme of the European Commission under FET-Open Grant No. 863155 (s-Nebula).
 The authors would like to thank the RENATECH network and acknowledge the support from the Government of the Czech Republic (doctoral grant competition CZ.02.2.69/0.0/0.0/19\_073/0016945 under the project DGS/TEAM/2020-027).
\end{acknowledgments}

\section*{Author Declarations}
The authors declare no competing interests.

\section*{Data Availability Statement}
The data that support the findings of this study are publicly available through the ZENODO server under the reference https://doi.org/10.5281/zenodo.5742851.

\section*{References}
%\nocite{*}
\bibliography{biblio}% Produces the bibliography via BibTeX.

\end{document}